\PassOptionsToPackage{table}{xcolor}
\documentclass[letterpaper, preprint]{article}
\usepackage{aaai2027}  
\usepackage[hyphens]{url}  
\usepackage{graphicx} 
\usepackage{natbib}  
\usepackage{caption} 
\usepackage{algorithm}
\usepackage{amsmath}  
\usepackage[table]{xcolor} 
\usepackage{newfloat}
\usepackage{listings}
\usepackage{multirow}
\DeclareCaptionStyle{ruled}{labelfont=normalfont,labelsep=colon,strut=off} 
\floatstyle{ruled}
\newfloat{listing}{tb}{lst}{}
\floatname{listing}{Listing}

\usepackage{booktabs}

\title{Selective KV Cache Protection for Noise-Resilient LLM Inference on Analog Compute-In-Memory Systems}
\author{
Yuannuo Feng\textsuperscript{\rm 1}\equalcontrib,
Wenyong Zhou\textsuperscript{\rm 2}\equalcontrib,
Yuang Ma\textsuperscript{\rm 3},
Yizhe Chen\textsuperscript{\rm 1},\\
Wenshuai Yao\textsuperscript{\rm 4},
Yuxin Xie\textsuperscript{\rm 1},
Ngai Wong\textsuperscript{\rm 2}\corresponding,
Wang Kang\textsuperscript{\rm 1}\corresponding
}
\affiliations{
\textsuperscript{\rm 1}School of Integrated Circuit Science and Engineering, Beihang University, Beijing, China\\
\textsuperscript{\rm 2}Department of Electrical and Computer Engineering, The University of Hong Kong, Hong Kong SAR, China\\
\textsuperscript{\rm 3}School of Microelectronic, University of Science and Technology of China, Hefei, China\\
\textsuperscript{\rm 4}School of Integrated Circuits, Peking University, Beijing, China
}

\begin{document}
\maketitle

\begin{abstract}

Analog compute-in-memory (CIM) arrays have emerged as a promising substrate for energy-efficient LLM inference, particularly for weight-stationary computations in linear layers. However, extending analog CIM to attention mechanisms introduces a fundamental challenge: KV cache operations demand repeated in-situ weight updates, and the resulting mismatch with the weight-stationary paradigm exposes dynamic computations to significant hardware noise, a critical problem that remains largely unexplored. 
In this paper, we present the first systematic study of dynamic attention computation on analog CIM arrays, revealing that initial and recent tokens exhibit disproportionate vulnerability to hardware noise. Motivated by this token-level insight, we propose a hierarchical token protection strategy that keeps sink tokens and a sliding recent-token window on a higher-precision digital path while processing the bulk KV cache on analog CIM. A co-designed scheduler combines analog programming, ownership transition, and bulk-MVM tile formation to bound digital overhead. 
Evaluations on nine LLMs show that our approach lowers average perplexity under analog noise from 33.91 to 11.95, close to the clean baseline of 11.06, while improving dynamic-KV programming-row utilization from 23.1\% to 91.2\%.
\end{abstract}




\section{Introduction}

\begin{figure}[!t]
\centering
\includegraphics[width=1.0\columnwidth]{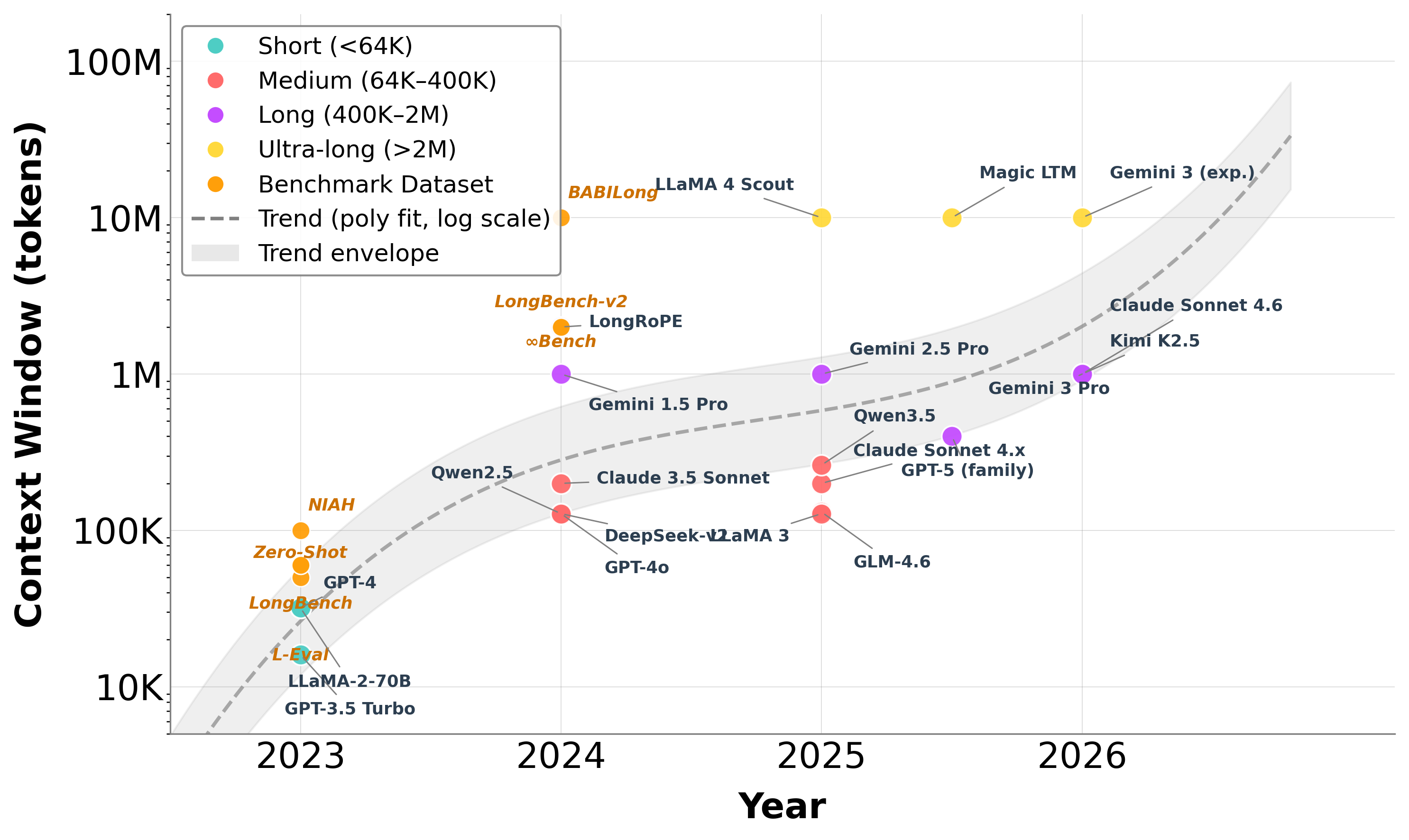}
\caption{The rapid expansion of LLM context windows from 2023 to 2026, growing from around 10K tokens to over 10M tokens.}
\label{fig:context_trend}
\end{figure}

Large language models (LLMs) have demonstrated remarkable capabilities across a wide range of tasks~\cite{Vaswani+2017,gpt3,gpt4,qwen}, driving an exponential expansion of their supported context windows. As illustrated in Figure~\ref{fig:context_trend}, the context window of state-of-the-art LLMs has grown from approximately 10K tokens in 2023 to over 10M tokens by 2026, a three-order-of-magnitude increase in just three years, with no sign of plateau~\cite{groeneveld2024olmo,qwen3}. This rapid scaling imposes severe computational demands: the self-attention mechanism exhibits quadratic complexity with respect to sequence length, making attention computation increasingly dominant in both latency and energy consumption for long-context inference~\cite{pope2023efficiently,sheng2023flexgen,vllm}.

Analog compute-in-memory (CIM) arrays have emerged as a compelling hardware substrate to address this challenge, offering orders-of-magnitude improvements in energy efficiency by performing matrix-vector multiplications (MVMs) directly within memory~\cite{verma2019memory,sebastian2020memory,ambrogio2018equivalent}. Existing analog CIM accelerators have demonstrated strong results for weight-stationary workloads, where model parameters are programmed once and reused across many inference steps. This pattern aligns naturally with the linear projection layers in Transformers~\cite{rramcim,sramcim,fefetcim,iccad_cim,iccad_cim_2}. However, the attention mechanism itself involves a fundamentally different computational pattern: the KV cache grows dynamically with context length and must be repeatedly read and updated, creating a significant mismatch with the weight-stationary paradigm of analog CIM~\cite{Vaswani+2017,guo2024cimformer}.

\begin{figure}[!t]
\centering
\includegraphics[width=1.0\columnwidth]{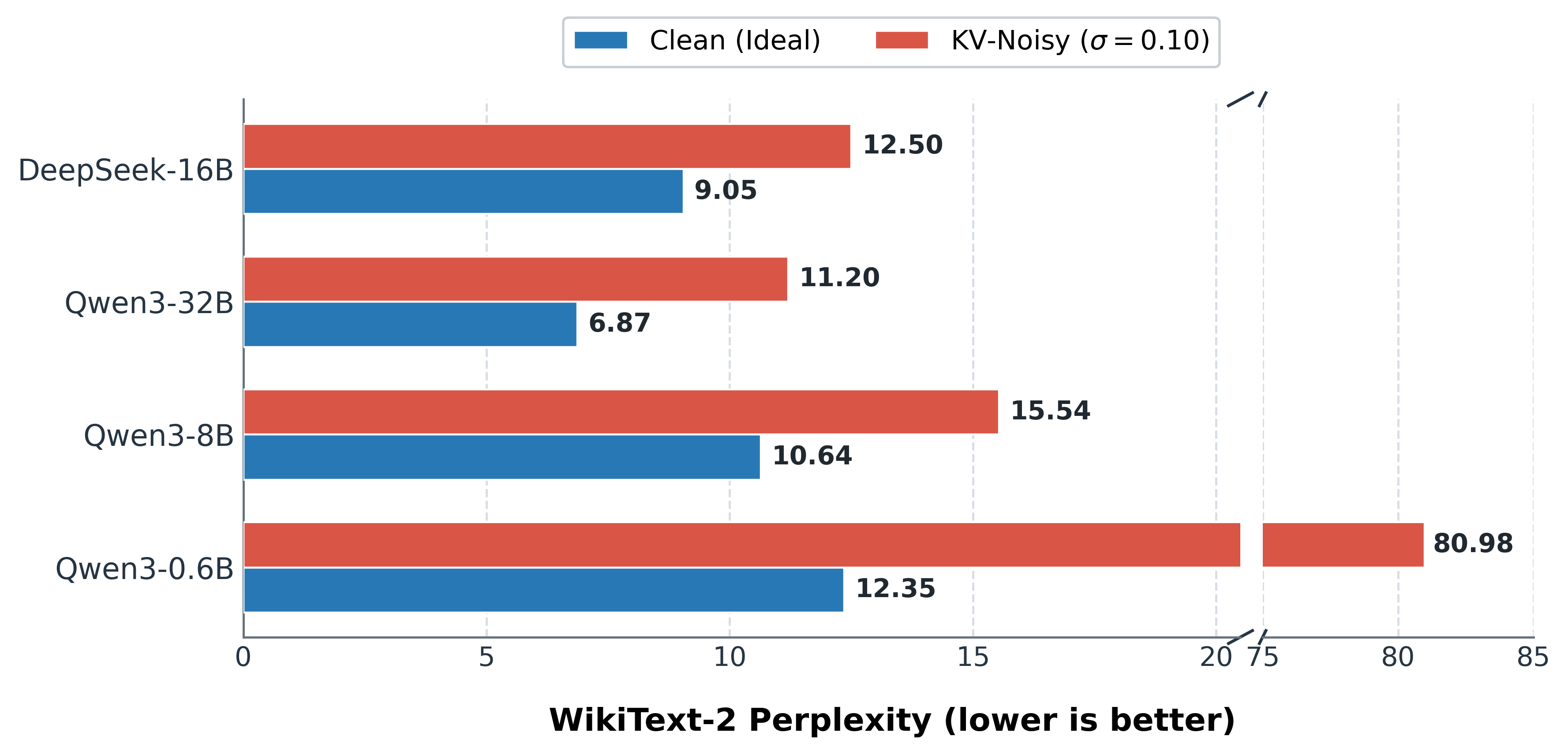}
\caption{Perplexity degradation caused by analog hardware noise injected into the KV cache across multiple LLMs.}
\label{fig:noise_impact}
\end{figure}

Recent advances in emerging non-volatile memory (NVMs) devices, such as resistive RAM (RRAM)~\cite{rramcim_dac} and phase-change memory (PCM), have enabled repeated in-situ weight programming at reasonable energy costs, making dynamic KV cache computation on analog CIM physically feasible. Nevertheless, a critical challenge remains largely overlooked: analog hardware is inherently susceptible to device-level noise (e.g., programming noise, read noise, and thermal fluctuations)~\cite{iccad_noise,iccad_noise_2}, and the robustness implications of performing dynamic attention computation under such noise conditions have not been systematically studied~\cite{CIM_noise,noiseguard,asicon1}. 
As shown in Figure~\ref{fig:noise_impact}, injecting noise into the KV cache sharply increases perplexity across multiple LLMs. For example, Qwen3-8B and Qwen3-14B increase from 10.64 to 21.54 and from 9.05 to 19.11, respectively, motivating explicit protection for dynamic analog KV storage.

Dynamic KV storage differs from static model-weight storage in both exposure and consequence. A static array holds one tensor that is reused across requests, so its characterization and compensation can be amortized. A KV entry, in contrast, is generated online, programmed during the current sequence, and immediately participates in many subsequent queries. Its useful lifetime and attention importance also depend on token position. Applying one uniform correction to this evolving state therefore ignores when an entry is written and how strongly later queries depend on it. Our central premise is that the hardware need not make every KV row equally precise: it should preserve the small position-defined subset whose corruption propagates most broadly, while using analog density for the remaining full sequence.

Prior work on analog CIM for LLM inference has predominantly focused on noise mitigation for static weight storage~\cite{asicon2,zhoudate,NAT,datecompensation}, or has addressed attention acceleration without accounting for hardware non-idealities~\cite{pope2023efficiently,guo2024cimformer}. To the best of our knowledge, no existing work provides a systematic analysis of noise vulnerability in dynamic attention computation on analog CIM, nor offers a hardware-algorithm co-design solution to address it.

In this paper, we present the first systematic investigation of dynamic attention computation on analog CIM arrays under noise. Our analysis reveals that \textit{initial tokens} and \textit{recent neighbor tokens} are disproportionately vulnerable to hardware noise. Building on this finding, we propose a hierarchical token protection framework and a co-designed configurable hardware architecture. 

The key contributions of this work are as follows:

\begin{itemize}
    \item We conduct the first systematic study of KV cache vulnerability under noise conditions, identifying asymmetric token-level vulnerability patterns.

    \item We propose an algorithm that selectively computes the most noise-sensitive tokens, including initial sink tokens and the sliding recent-token window, on a higher-precision digital path while retaining the bulk KV cache on analog CIM.

    \item We design a flexible CIM architecture that supports seamless switching between prefill-only and prefill-decode modes to enable various scenarios.

    \item Experiments show that GoS lowers average perplexity under analog noise from 33.91 to 11.95 and raises dynamic-KV programming-row utilization from 23.1\% to 91.2\%.
\end{itemize}

\section{Related Work}
\label{sec:background}

Transformer attention computes~\cite{Vaswani+2017}
\begin{equation}
    \text{Attention}(\mathbf{Q}, \mathbf{K}, \mathbf{V}) =
    \text{softmax}\!\left(
    \frac{\mathbf{Q}\mathbf{K}^{\top}}{\sqrt{d_k}}
    \right)\mathbf{V},
    \label{eq:attention}
\end{equation}
where $\mathbf{Q}$, $\mathbf{K}$, and $\mathbf{V}$ are generated by learned projections. During autoregressive decoding, previous keys and values are stored in the KV cache, and the current query attends to all cached entries:
\begin{equation}
    \mathbf{o}_t =
    \text{softmax}\!\left(
    \frac{\mathbf{q}_t\mathbf{K}_{1:t}^{\top}}{\sqrt{d_k}}
    \right)\mathbf{V}_{1:t}.
    \label{eq:kvcache}
\end{equation}
With rapidly increasing context length, KV-cache storage and attention computation have become major bottlenecks. Existing methods reduce this overhead through token eviction~\cite{h2o,liu2023scissorhands}, KV quantization~\cite{kvcache_quant,kvquant}, sparse attention~\cite{longformer}, and model-level designs such as GQA~\cite{gqa} and MLA~\cite{deepseek}. These approaches mainly optimize software memory usage and do not address the implications of executing KV-cache attention on analog hardware.

Analog CIM performs MVM in memory by encoding matrix entries as conductance values:
\begin{equation}
    \mathbf{I} = \mathbf{G}\mathbf{V}_{\text{in}},
    \label{eq:cim_mvm}
\end{equation}
where $\mathbf{G}$ is the stored matrix and $\mathbf{I}$ is digitized by ADCs. Existing analog CIM accelerators for LLMs largely assume a \textit{weight-stationary} execution model~\cite{2,3}, which is suitable for static projection and FFN weights. Some recent works extend CIM to attention projection layers~\cite{4,5}, but they do not address the dynamic KV cache, which is generated online and updated at every decoding step.

Prior robustness techniques, including noise-aware training~\cite{NAT}, quantization~\cite{kvcache_quant}, and error correction~\cite{2}, mainly target static weight storage. In contrast, dynamic KV-cache entries are repeatedly programmed and read, making them exposed to analog programming and read noise. This work studies this underexplored setting and identifies token-level vulnerability patterns that enable selective hybrid analog--digital protection.

\section{Methodology}

\subsection{Analysis of KV Cache Under Analog Noise}
Attention exhibits the well-known attention-sink phenomenon~\cite{streaming}: the \texttt{<BOS>} token attracts disproportionately large attention across positions and layers. Since the softmax operator must allocate probability mass even when no key is strongly relevant, models often route excess mass to this semantically neutral token. Consequently, corruption of the \texttt{<BOS>} key can affect all queries globally.

\begin{figure}[!t]
\centering
\includegraphics[width=0.95\columnwidth]{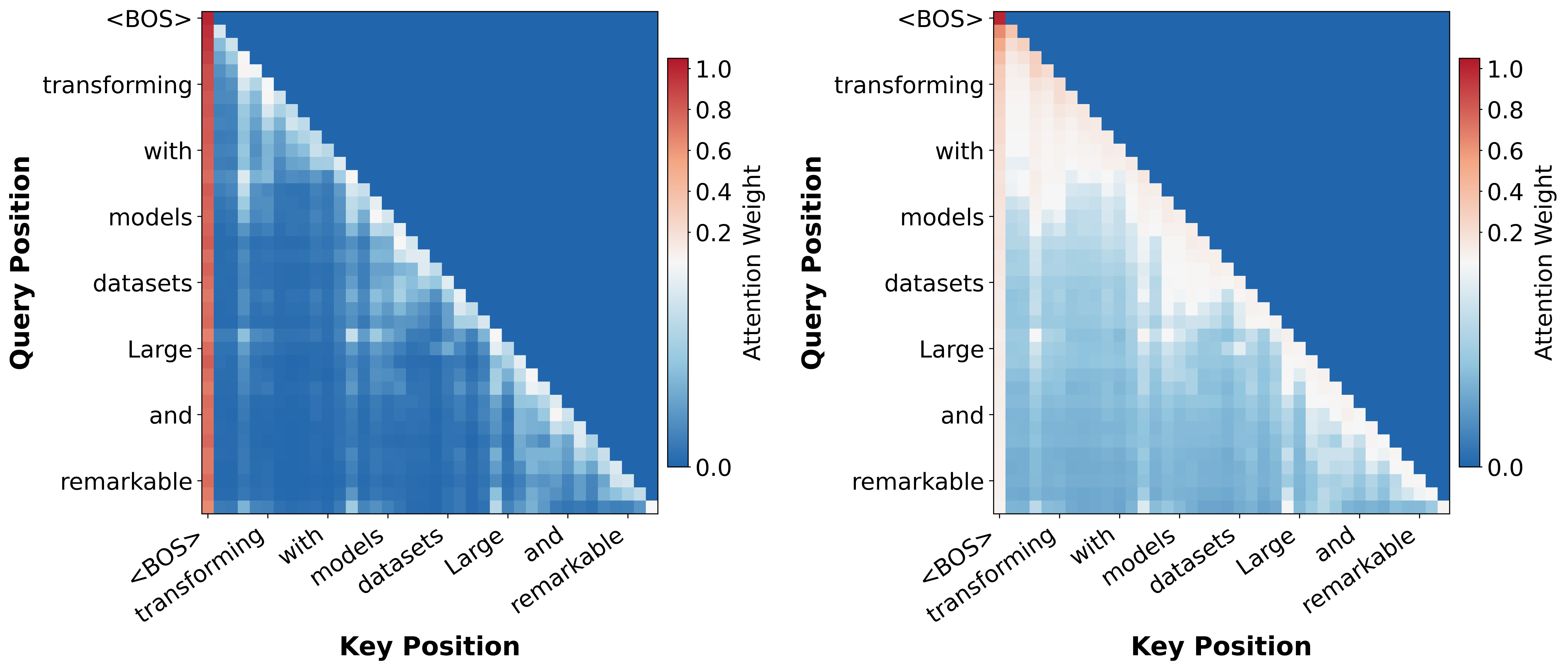}
\caption{Left: Averaged attention map of Qwen3-8B under clean KV cache. Right: Attention map under KV noise.}
\label{fig:attention_map}
\end{figure}

Figure~\ref{fig:attention_map} illustrates this vulnerability. Clean attention shows strong \texttt{<BOS>} dominance and structured diagonal patterns, whereas KV-noisy attention collapses toward a flatter distribution because analog noise randomizes key vectors and weakens pre-softmax score contrast. This reveals three sensitivity tiers: sink tokens, whose corruption has global impact; recent tokens, which carry local context and receive high attention; and middle bulk tokens, whose low individual attention makes them relatively noise-tolerant and suitable for analog processing.

\begin{figure}[!t]
\centering
\includegraphics[width=0.95\columnwidth]{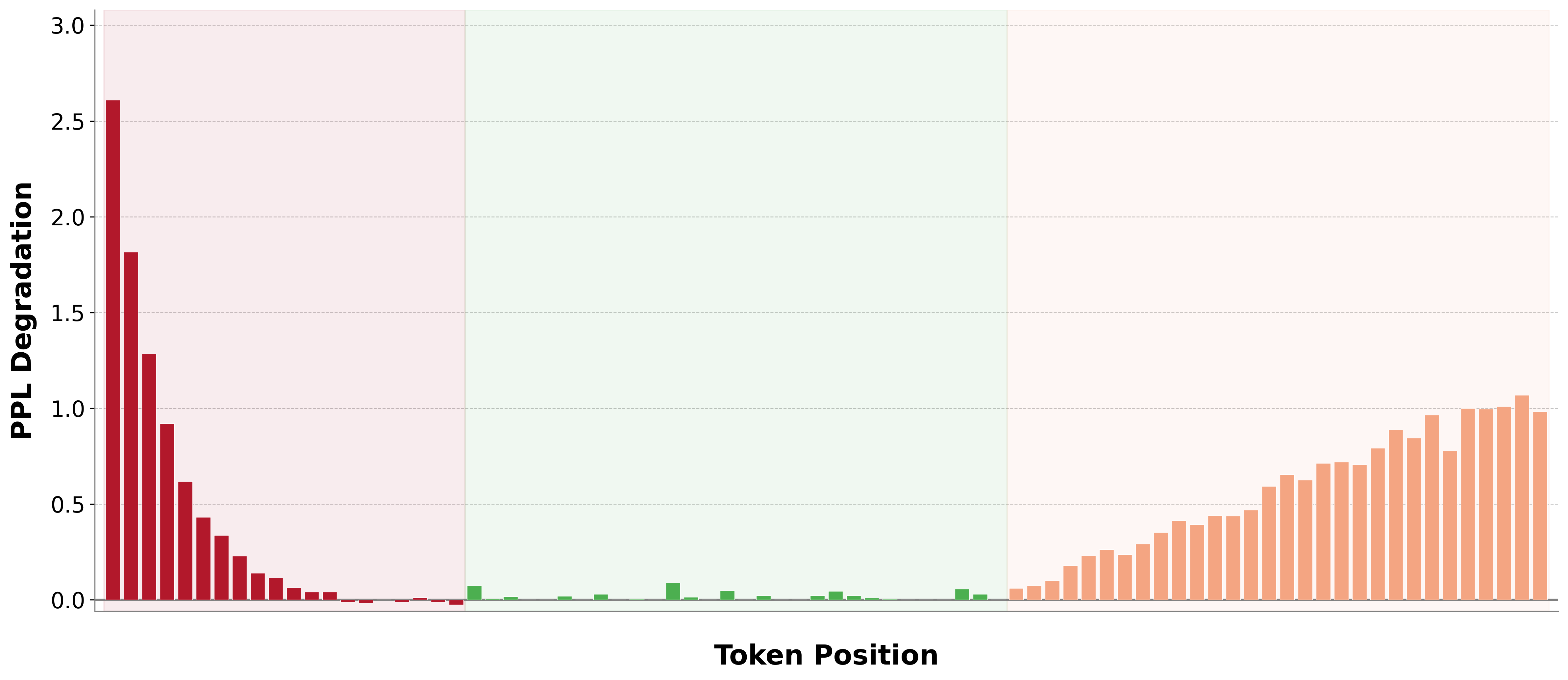}
\caption{Per-position PPL degradation of Qwen3-8B from analog hardware noise across token sequence.}
\label{fig:position_ppl}
\end{figure}

Figure~\ref{fig:position_ppl} further quantifies this asymmetry through per-position perplexity degradation. Head tokens suffer the largest degradation, while middle tokens show near-zero degradation. Tail tokens become increasingly sensitive near recent positions due to concentrated local-context information. These results motivate GoS: protect a small set of sink and recent tokens with digital precision while mapping the noise-tolerant middle bulk to analog CIM.

\subsection{Guard-of-Sink Token Partitioning}
GoS exploits the asymmetric token vulnerability identified above by assigning only noise-sensitive KV entries to a higher-precision digital path. At decoding step $t$, the logical KV cache contains token positions $\{1,\ldots,t\}$. GoS partitions the cache into a protected set and an analog bulk set. The protected set consists of a pinned sink region and a sliding recent region:
\begin{equation}
\mathcal{S}_{\mathrm{sink}}(t)
=
\{1,\ldots,\min(m,t)\},
\end{equation}
\begin{equation}
\mathcal{S}_{\mathrm{recent}}(t)
=
\{\max(m+1,t-r+1),\ldots,t\},
\end{equation}
\begin{equation}
\mathcal{S}_{\mathrm{GoS}}(t)
=
\mathcal{S}_{\mathrm{sink}}(t)
\cup
\mathcal{S}_{\mathrm{recent}}(t),
\qquad
r=k-m,
\end{equation}
where $m$ is the number of pinned sink tokens, $k$ is the total protection budget, and $r$ is the recent-window size. The remaining positions form the analog bulk set:
\begin{equation}
\mathcal{S}_{\mathrm{bulk}}(t)
=
\{1,\ldots,t\}
\setminus
\mathcal{S}_{\mathrm{GoS}}(t).
\end{equation}
Let $\mathcal{S}_{\mathrm{pending}}(t)\subseteq\mathcal{S}_{\mathrm{bulk}}(t)$ denote entries awaiting batched analog programming. The physical path for token position $i$ is
\begin{equation}
\mathrm{path}(i,t)
=
\begin{cases}
\mathrm{Digital}, & i\in \mathcal{S}_{\mathrm{GoS}}(t)\cup\mathcal{S}_{\mathrm{pending}}(t),\\
\mathrm{Analog\ CIM}, & i\in \mathcal{S}_{\mathrm{bulk}}(t)\setminus\mathcal{S}_{\mathrm{pending}}(t).
\end{cases}
\label{eq:gos_path}
\end{equation}
This set-based definition naturally handles early decoding steps, where the sink and recent regions may overlap and fewer than $k$ tokens are protected.

GoS does not recompute protected tokens using a clean digital projection. The Q/K/V projections are still produced by the analog projection arrays. The digital path only stores and reads a higher-precision copy of selected K/V entries, thereby bypassing dynamic analog-KV programming noise and read noise for these entries while retaining any projection noise already present in the generated K/V vectors.

The partition also separates protection from sparsification. Every position remains present in the score vector: GoS changes the storage and execution path of a token but neither evicts it nor approximates attention with a truncated key set. This distinction matters for long contexts, where a token with low instantaneous attention may become relevant to a later query. The sink budget $m$ is pinned because its influence is global, whereas the remaining $k-m$ slots rotate because recency is time dependent. Consequently, the policy requires only token indices and fixed deployment parameters; it does not introduce a learned predictor, request-specific retraining, or an auxiliary importance pass.

\begin{figure}[!t]
\centering
\includegraphics[width=1.0\columnwidth]{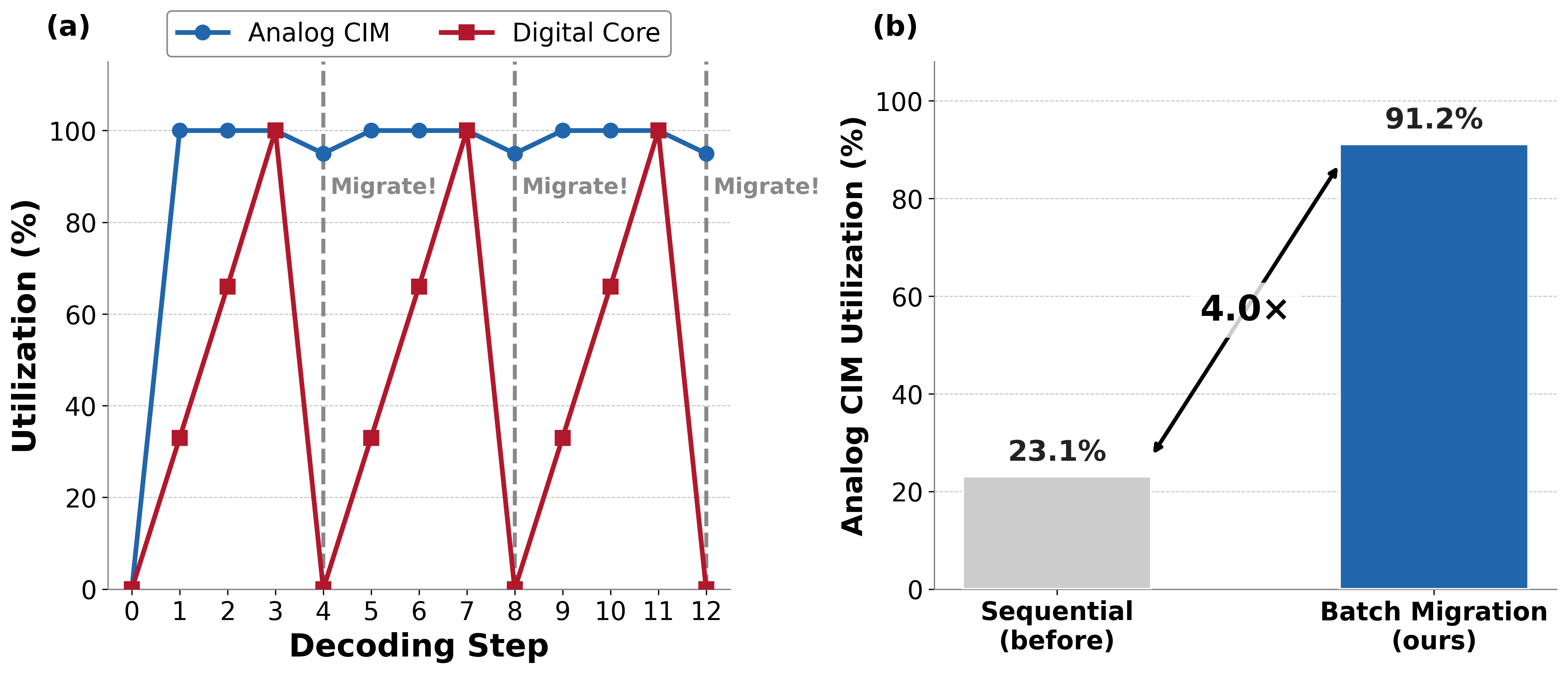}
\caption{Dynamic-KV programming-row utilization under sequential and threshold-coalesced scheduling.}
\label{fig:tile_util}
\end{figure}
\begin{figure*}[t]
\centering
\includegraphics[width=0.95\textwidth]{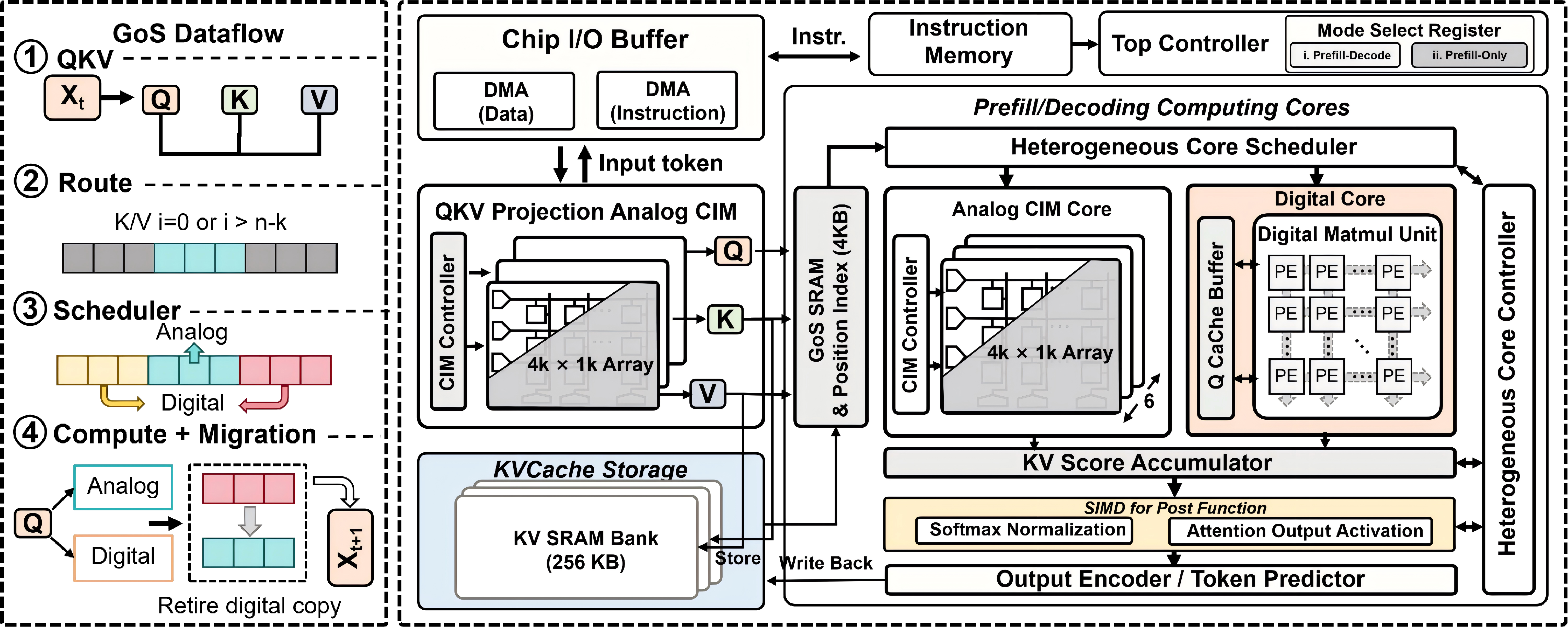}
\caption{GoS-enabled analog CIM accelerator architecture with hybrid analog-digital processing and 3D-integrated design.}
\label{fig:architecture}
\end{figure*}

The hybrid attention computation uses one global softmax over all token positions. For each position $i$, the pre-softmax score is
\begin{equation}
s_t[i]
=
\begin{cases}
\dfrac{\widehat{\mathbf q}_t (\mathbf k_i^{\mathrm{dig}})^{\top}}{\sqrt{d_k}},
&
i\in \mathcal{S}_{\mathrm{GoS}}(t)\cup\mathcal{S}_{\mathrm{pending}}(t),\\[6pt]
\dfrac{\widehat{\mathbf q}_t \widetilde{\mathbf k}_i^{\top}}{\sqrt{d_k}},
&
i\in \mathcal{S}_{\mathrm{bulk}}(t)\setminus\mathcal{S}_{\mathrm{pending}}(t),
\end{cases}
\label{eq:hybrid_score}
\end{equation}
where $\mathbf k_i^{\mathrm{dig}}$ is the higher-precision digital key and $\widetilde{\mathbf k}_i$ is the analog-resident key affected by dynamic KV-storage noise. The complete score vector is then normalized as
\begin{equation}
\alpha_t = \mathrm{softmax}(s_t),
\end{equation}
and the attention output is computed as
\begin{equation}
o_t
=
\sum_{i\in \mathcal{S}_{\mathrm{GoS}}(t)\cup\mathcal{S}_{\mathrm{pending}}(t)}
\alpha_t[i] v_i^{\mathrm{dig}}
+
\sum_{i\in \mathcal{S}_{\mathrm{bulk}}(t)\setminus\mathcal{S}_{\mathrm{pending}}(t)}
\alpha_t[i] \widetilde{v}_i,
\label{eq:hybrid_value}
\end{equation}
where $v_i^{\mathrm{dig}}$ and $\widetilde{v}_i$ denote the digital and analog-resident value vectors, respectively. This formulation preserves full-attention semantics during migration: every token is served by exactly one path, and normalization is performed once globally after positional merging.

\subsection{Coalesced Migration and Tile Scheduling}
The token partition in Eq.~\eqref{eq:gos_path} determines the logical ownership of each KV entry, but an efficient analog implementation also requires dense tile formation. A naive implementation would immediately migrate one token from the recent window to the analog bulk region at every decoding step. Since analog CIM macros are row-parallel, such token-by-token services activate only a small fraction of the available rows, leading to poor utilization and frequent digital--analog synchronization.

To avoid this inefficiency, GoS decouples logical ownership transition from physical tile activation. Each KV entry follows a three-state ownership protocol:
\begin{itemize}
    \item \textbf{Protected}: the token belongs to $\mathcal{S}_{\mathrm{GoS}}(t)$ and has a valid higher-precision digital copy.
    \item \textbf{Pending-bulk}: the token has logically left the recent window but retains a valid digital copy and continues to participate through the digital path while awaiting batched analog programming.
    \item \textbf{Active-bulk}: the token's analog row has been programmed and verified; ownership has switched atomically to the analog path and the temporary digital copy can be reclaimed.
\end{itemize}

At each decoding step, the newly generated K/V pair is inserted into the recent digital buffer. When the recent boundary advances, a departing token that is not pinned as a sink enters the pending buffer but remains digitally accessible. Once the pending population reaches the migration threshold $\theta$, the scheduler programs and verifies the K/V rows as one dense analog batch. It then atomically changes their ownership to active-bulk before reclaiming the digital copies. Consequently, no token is omitted from attention during migration, and the temporary pending storage is bounded by $\theta-1$ entries, giving a maximum digital footprint of $k+\theta-1$ KV entries.

Correctness follows from two ownership invariants. First, a token has exactly one serving path at each attention operation: protected and pending entries are read digitally, while only verified active-bulk entries are read from the analog bank. Second, reclamation is ordered after the metadata commit, so a failed or incomplete programming service leaves the digital copy authoritative. Coalescing therefore changes when physical rows become active, not the logical sequence seen by attention. The threshold trades temporary digital capacity against row occupancy, but it does not alter the protected set or the attention equation. These invariants also make storage scale independently of context length: the protected circular buffer is fixed at $k$, and unfinished migrations contribute fewer than $\theta$ additional entries.

We define the dynamic-KV programming-row utilization as
\begin{equation}
U_{\mathrm{prog}}
=
\frac{\sum_{b=1}^{B} n_b}
{B R_{\mathrm{tile}}},
\label{eq:cim_util}
\end{equation}
where $n_b$ is the number of valid rows written by programming service $b$, $R_{\mathrm{tile}}$ is the row capacity of one tile service, and $B$ is the number of programming services in the measurement window. Figure~\ref{fig:tile_util} illustrates the effect of coalescing. Without coalescing, sequential token migration produces sparse programming services and low valid-row occupancy. With threshold-coalesced scheduling, pending rows are programmed as dense batches, substantially improving programming-row utilization. This metric characterizes dynamic-KV programming efficiency rather than attention-compute utilization.
 
\subsection{Hardware Architecture}
Figure~\ref{fig:architecture} maps the GoS dataflow to a hybrid analog--digital CIM accelerator. The architecture implements three mechanisms required by Eqs.~\eqref{eq:gos_path}--\eqref{eq:hybrid_value}: ownership metadata for token routing, dual-path score generation, and positional score merging followed by softmax.

\textbf{QKV projection stage.}
The static projection matrices $W_Q$, $W_K$, and $W_V$ are stored in weight-stationary analog CIM arrays. For each input token, the projection arrays generate $q_t$, $k_t$, and $v_t$ through analog MVMs. These arrays are programmed once at model deployment and remain stationary during inference, so they do not incur dynamic KV-cache reprogramming overhead.

\textbf{Protected and bulk KV storage.}
The protected SRAM is logically divided into an $m$-entry pinned sink buffer, an $r$-entry circular recent buffer, and a temporary pending buffer of at most $\theta-1$ entries. The sink buffer is filled during the first $m$ decoding steps and is never evicted. The recent buffer is updated every step and retains the most recent $r=k-m$ KV entries. Pending entries remain on the digital path until they are programmed into the analog KV bank as a dense batch. The dynamic KV bank is logically split into a K bank for score computation and a V bank for value aggregation, although the two banks may be physically separated or time-multiplexed depending on the macro organization.

\textbf{Ownership scheduler and tile formation.}
The heterogeneous scheduler maintains per-token ownership metadata, indicating whether each token is protected, pending-bulk, or active-bulk. It evaluates Eq.~\eqref{eq:gos_path}, retains pending digital entries until a batched analog write has been verified, and then atomically switches their ownership according to the threshold policy described in Section~3.3. This scheduler is the hardware mechanism behind the utilization improvement shown in Figure~\ref{fig:tile_util}.

\textbf{Dual-path attention execution.}
During attention computation, the digital PE array computes scores for protected and pending tokens, while the analog CIM core computes scores for active-bulk tokens. The KV Score Accumulator merges the two score streams according to their original token positions and forwards the complete score vector to the softmax unit. The resulting attention weights are then split back to the two paths for value aggregation, and the partial outputs are accumulated to produce the final attention output. Since both paths share one global softmax and ownership switches atomically, the computation remains mathematically equivalent to full attention except for analog noise affecting active-bulk entries.

\textbf{Operating modes.}
In \emph{Prefill-Only}, analog prefill retains a complete digital KV copy in off-chip HBM and stages it through SRAM for digital decode; the reported cost includes this transfer. In \emph{Prefill-Decode}, analog CIM serves active-bulk entries while digital PEs serve sink, recent, and pending entries, bounding on-chip digital storage by $k+\theta-1$.

\section{Experiments}
\label{sec:experiments}
\subsection{Experiment Setup}

We evaluate Qwen3, Llama, DeepSeek, OLMo, and OLMoE models on WikiText-2, ARC-Challenge, PIQA, GSM8K, and MATH500. We use each benchmark's standard accuracy protocol and the same prompting and decoding settings for clean and noisy runs. Unless otherwise stated, $m=8$, $k=128$, and the dynamic-KV working set is 2.56K tokens; the resulting 5\% protection ratio is relative to that working set (1.56\% for an 8K full sequence).

\begin{figure}[!t]
\centering
\includegraphics[width=\columnwidth]{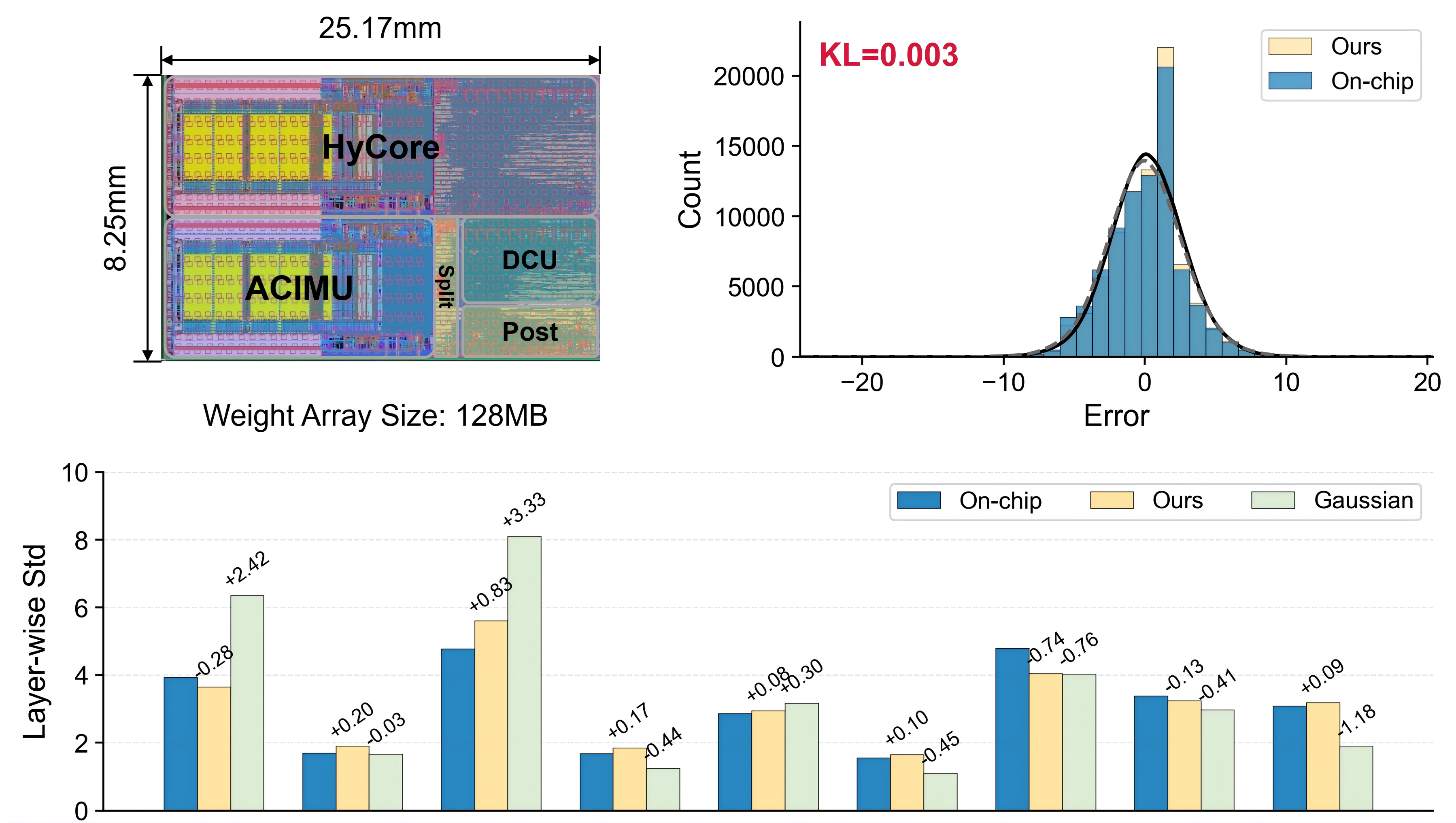}
\caption{Measured-chip calibration of the configurable CIM noise model; held-out validation yields a KL divergence of 0.003.}
\label{fig:measured_noise_model}
\end{figure}

Rather than assuming an idealized Gaussian perturbation, we derive the noise kernel from measurements of hundreds of CIM chips represented by the prototype in Figure~\ref{fig:measured_noise_model}. Data-driven fitting extracts the measured error distribution, while configurable device, programming, readout, and conversion hyperparameters allow the model to cover different CIM operating conditions. Validation against held-out on-chip measurements yields a KL divergence of only 0.003, supporting its use as a generalizable noise source in the following experiments.

All comparisons at a given operating point share the checkpoint, input sequence, protection budget, and evaluation metric. Clean inference removes the modeled analog non-idealities; the unprotected analog baseline exposes all KV entries to the dynamic-storage path; and GoS changes only the ownership of the selected entries. The two GoS modes use the same token policy but differ in where bulk decode executes, so their quality and cost should be read jointly rather than as interchangeable configurations. The supplementary material gives the complete noise equations, representative memory organization, migration assumptions, baseline adaptations, and per-layer cost accounting.

\begin{table*}[t]
\centering
\small
\setlength{\tabcolsep}{2pt}
\begin{tabular}{lcccccccccc}
\toprule
\multirow{2}{*}{\textbf{Method}}
& \textbf{Qwen3} & \textbf{Qwen3} & \textbf{Qwen3} & \textbf{Llama3.2} & \textbf{Llama3.2} & \textbf{Llama3} & \textbf{DeepSeek} & \textbf{OLMO} & \textbf{OLMOE} & \multirow{2}{*}{\textbf{AVG}} \\
& \textbf{0.6B} & \textbf{8B} & \textbf{32B} & \textbf{1B} & \textbf{3B} & \textbf{8B} & \textbf{16B} & \textbf{7B} & \textbf{7B} & \\
\midrule
Clean (Ideal) & 12.35 & 10.64 & 6.87 & 15.42 & 12.79 & 9.26 & 9.05 & 16.11 & 7.02 & 11.06 \\
Analog KV Cache & 80.98 & 15.54 & 11.20 & 100.78 & 38.84 & 19.00 & 12.50 & 17.33 & 9.02 & 33.91 \\
k-b Calibration & 70.66 & 11.53 & 8.19 & 63.34 & 38.71 & 18.95 & 12.49 & 17.32 & 9.01 & 27.80 \\
Averaging & 18.05 & 10.76 & 7.80 & 19.73 & 14.60 & 9.75 & 10.50 & 16.46 & 7.70 & 12.82 \\
\rowcolor{blue!15}
\textbf{Ours (Prefill-Decode)} & \textbf{14.92} & \textbf{10.67} & \textbf{7.38} & \textbf{17.85} & \textbf{13.68} & \textbf{9.53} & \textbf{9.72} & \textbf{16.32} & \textbf{7.45} & \textbf{11.95} \\
\rowcolor{orange!15}
\textbf{Ours (Prefill-Only)} & \textbf{13.58} & \textbf{10.65} & \textbf{7.15} & \textbf{16.73} & \textbf{13.24} & \textbf{9.42} & \textbf{9.38} & \textbf{16.18} & \textbf{7.28} & \textbf{11.51} \\
\bottomrule
\end{tabular}
\caption{WikiText-2 perplexity under analog KV noise (8K context, $m=8$, $k=128$).}
\label{tab:cim_noise_results}
\end{table*}

\subsection{Experiment Results}

\begin{figure}[!t]
\centering
\includegraphics[width=\columnwidth]{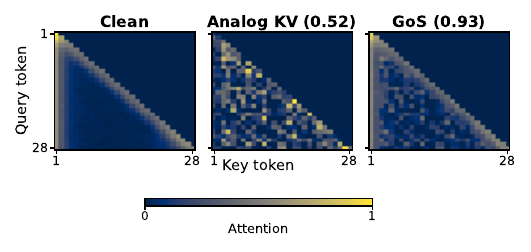}
\caption{Representative attention case before and after GoS protection; parenthesized values denote cosine similarity to the clean map.}
\label{fig:restoration_case}
\end{figure}

Figure~\ref{fig:restoration_case} provides a qualitative, representative simulation of the mechanism. Analog KV noise disperses the causal attention structure and suppresses the sink region, reducing cosine similarity to the clean attention map from 1.00 to 0.52. Protecting sink and recent entries restores the dominant structure and raises similarity to 0.93; residual differences remain in the unprotected analog bulk.

We separate projection noise from dynamic-KV storage noise. Analog Q/K/V projections first produce
\begin{align}
\widehat{\mathbf q}_t &= \widetilde{\Phi}_{\mathrm{CIM}}(\widetilde{\mathbf W}_Q,\mathbf x_t), \nonumber\\
\widehat{\mathbf k}_t &= \widetilde{\Phi}_{\mathrm{CIM}}(\widetilde{\mathbf W}_K,\mathbf x_t), \nonumber\\
\widehat{\mathbf v}_t &= \widetilde{\Phi}_{\mathrm{CIM}}(\widetilde{\mathbf W}_V,\mathbf x_t),
\end{align}
where $\widetilde{\Phi}_{\mathrm{CIM}}$ includes projection-array non-idealities. When an entry is programmed into and read from the analog KV bank, its stored representations become
\begin{align}
\widetilde{\mathbf k}_i &= \widehat{\mathbf k}_i
+\boldsymbol\epsilon^{K}_{i,\mathrm{prog}}
+\boldsymbol\epsilon^{K}_{i,\mathrm{read}}, \nonumber\\
\widetilde{\mathbf v}_i &= \widehat{\mathbf v}_i
+\boldsymbol\epsilon^{V}_{i,\mathrm{prog}}
+\boldsymbol\epsilon^{V}_{i,\mathrm{read}}.
\end{align}
Protected and pending entries instead use $\mathbf k_i^{\mathrm{dig}}=\widehat{\mathbf k}_i$ and $\mathbf v_i^{\mathrm{dig}}=\widehat{\mathbf v}_i$. Thus, digital protection bypasses dynamic programming and read noise but is not credited with removing projection error. We use 8-bit weights, activations, DAC, and ADC, with default severity $\sigma=0.10$.

\textbf{Model Quality Evaluation.}
Table~\ref{tab:cim_noise_results} compares WikiText-2 results across nine models under clean inference, unprotected analog execution, and two representative mitigation baselines. Unprotected KV noise is particularly harmful to small models: PPL increases from 12.35 to 80.98 for Qwen3 (0.6B) and from 15.42 to 100.78 for Llama3.2 (1B). GoS reduces average PPL from 33.91 under unprotected analog inference to 11.95 in Prefill-Decode and 11.51 in Prefill-Only, compared with 11.06 for clean inference. For Qwen3 (0.6B), the corresponding values are 14.92 and 13.58, showing that digital decode fallback improves quality at the cost reported in Table~\ref{tab:hardware_eval}.

Performance variation across models is also important. The unprotected penalty ranges from modest degradation on OLMo and OLMoE to severe collapse on the two smallest dense models. Nevertheless, Prefill-Decode remains close to each model's clean reference across architecture families. This agrees with the positional design of GoS, which does not depend on hidden width, KV-head count, or MoE routing rules. Remaining gaps arise from analog bulk-storage and projection noise that selective KV protection cannot eliminate, explaining why Prefill-Only is generally, though not always, closer to clean inference.

\begin{figure}[!t]
\centering
\includegraphics[width=\columnwidth]{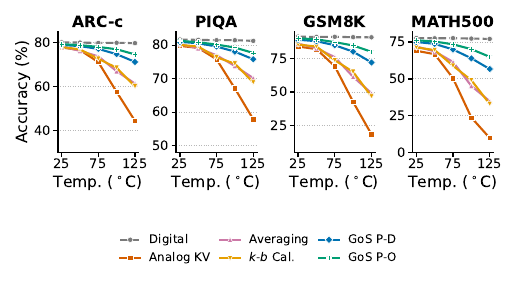}
\caption{Temperature-dependent Qwen3-8B accuracy on four downstream tasks under CIM noise.}
\label{fig:downstream}
\end{figure}

\textbf{Downstream Task Evaluation.}
For the temperature sweep in Figure~\ref{fig:downstream}, the chip-calibrated model maps 25, 50, 75, 100, and 125$^\circ$C to $\sigma=0.10$, 0.2, 0.3, 0.4, and 0.5, respectively. The clean digital reference bypasses analog non-idealities. All methods use identical Qwen3-8B checkpoints, prompts, and decoding settings at each operating point.

Figure~\ref{fig:downstream} evaluates robustness across four reasoning and language-understanding tasks. Unprotected analog KV inference changes little at low temperatures but degrades sharply in the high-noise regime, with the largest losses on GSM8K and MATH500. Averaging and $k$-$b$ calibration delay this degradation but yield inconsistent gains across tasks and temperatures. Both GoS modes remain close to the clean digital reference throughout the sweep. Prefill-Only achieves the highest accuracy because decode remains digital, whereas Prefill-Decode retains a clear advantage over the adapted baselines while preserving analog decode efficiency.

\textbf{Hardware Efficiency Evaluation.}
Table~\ref{tab:hardware_eval} reports the decode tradeoff at $100^\circ$C ($\sigma=0.40$), with energy and latency normalized to unprotected analog KV inference. Prefill-Decode adds 3\% energy and 4\% latency while achieving 80.2\% GSM8K accuracy. Prefill-Only uses digital decode and reaches 84.8\% accuracy at higher cost. All estimates use the same $m=8$, $k=128$ configuration.

\begin{figure}[!t]
\centering
\includegraphics[width=\columnwidth]{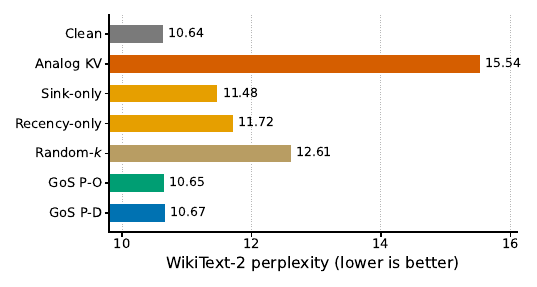}
\caption{Token-selection ablation on WikiText-2 at 8K context.}
\label{fig:token_ablation}
\end{figure}

\textbf{Token Protection Strategies.}
Figure~\ref{fig:token_ablation} isolates the selection policy at an 8K context length. The comparison includes clean storage, unprotected analog KV storage, and several selective policies: Sink-only (11.48 PPL), Recency-only (11.72), and Random-$k$ (12.61). Jointly protecting sink and recent tokens reduces PPL to 10.67 in Prefill-Decode, while Prefill-Only reaches 10.65 by eliminating analog bulk decode. These results show that the fixed budget protects two complementary sensitivity tiers rather than merely reducing the number of analog rows.

\begin{figure}[!t]
\centering
\includegraphics[width=\columnwidth]{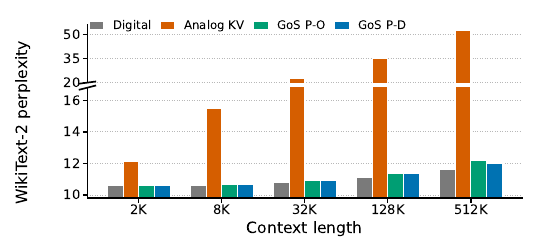}
\caption{Context-length scaling on WikiText-2 with a fixed GoS protection budget.}
\label{fig:context_ablation}
\end{figure}

\textbf{Context-Length Scaling.}
Figure~\ref{fig:context_ablation} extends WikiText-2 evaluation from 2K to 512K tokens. The broken vertical axis separates near-clean results from the much larger unprotected-analog values. Unprotected analog PPL increases from 12.08 to 52.63, whereas Prefill-Decode reaches 12.02 at 512K, only 0.46 above clean and 0.16 below Prefill-Only. At 512K, the cache is organized as 64 cold-storage tiles of 8K tokens, totaling approximately 32 GB, and streamed through HBM/SRAM staging. The on-chip protected-plus-pending footprint remains bounded by $k+\theta-1$ entries. Context growth therefore increases cold-tile traffic rather than the bounded on-chip digital footprint.

\begin{table}[!t]
\centering
\small
\setlength{\tabcolsep}{3pt}
\begin{tabular}{lccc}
\toprule
\textbf{Method} &
\textbf{Energy $\downarrow$} &
\textbf{Latency $\downarrow$} &
\textbf{Accuracy $\uparrow$} \\
\midrule
Analog KV Cache       & 1.00$\times$ & 1.00$\times$ & 42.7\% \\
Averaging              & 1.30$\times$ & 1.35$\times$ & 61.4\% \\
$k$-$b$ Calibration   & 1.15$\times$ & 1.16$\times$ & 65.1\% \\
\midrule
GoS (Prefill-Only)    & 1.16$\times$ & 1.14$\times$ & \textbf{84.8\%} \\
GoS (Prefill-Decode)  & \textbf{1.03$\times$} & \textbf{1.04$\times$} & 80.2\% \\
\bottomrule
\end{tabular}
\caption{Normalized decode cost and GSM8K accuracy at $100^\circ$C ($\sigma=0.40$, $m=8$, $k=128$).}
\label{tab:hardware_eval}
\end{table}

\subsection{Scope and Limitations}
The measured chip data calibrate the noise kernel, but the evaluated LLMs are not executed end to end on that prototype. The model captures the measured error distribution and held-out agreement. The default $m$, $k$, and migration threshold are fixed for the reported configuration; their best values may change with model architecture, device characteristics, or service-level constraints. GoS protects dynamic KV storage but deliberately retains projection-array error, and its bounded on-chip digital footprint does not remove HBM traffic for cold long-context tiles. Finally, the reported energy, latency, and PPA values are design estimates under the stated organization rather than measurements from a fabricated full-system GoS accelerator.

\section{Conclusion}

GoS is a co-designed analog CIM architecture that improves the noise robustness of dynamic KV cache computation in LLM inference by selectively protecting noise-sensitive tokens with a bounded digital buffer while storing noise-tolerant tokens on analog CIM arrays. By exploiting the asymmetric vulnerability of sink and recent tokens, GoS enables robust hybrid analog-digital attention with low protection overhead. Across nine LLMs, GoS reduces average noisy PPL from 33.91 to 11.95, approaching the clean baseline of 11.06, and increases dynamic-KV programming-row utilization from 23.1\% to 91.2\%.



\bibliography{aaai2027}
\end{document}